\documentstyle[aps,prl,twocolumn,epsf]{revtex}
\begin{document}
\draft
\wideabs{
\title{Weakly bound dimers of fermionic atoms}  
\author{D.S. Petrov$^{1,2}$, C. Salomon$^{3}$, and G.V.
Shlyapnikov$^{1,2,3}$}
\address{${}^1$ FOM Institute for Atomic and Molecular
Physics, Kruislaan 407,
1098 SJ Amsterdam, The Netherlands \\
${}^2$ Russian Research Center Kurchatov Institute, 
Kurchatov Square, 123182 Moscow, Russia \\ 
${}^3$ Laboratoire Kastler Brossel, 24 rue
Lhomond,
F-75231 Paris Cedex 05, France} 
\date{\today}
\maketitle
\begin{abstract}
We discuss the behavior of weakly bound bosonic dimers 
formed in a cold Fermi gas at a large positive scattering 
length $a$ for the interspecies interaction. We find the 
exact solution for the dimer-dimer elastic scattering and 
obtain a strong decrease of their collisional relaxation 
and decay with increasing $a$. The large ratio of the 
elastic to inelastic rate is promising for achieving 
Bose-Einstein condensation of the dimers and cooling the 
condensed gas to very low temperatures.   
\end{abstract}
}
The last years were marked by remarkable achievements in the
physics
of cold Fermi gases. Several groups succeeded in cooling
trapped fermionic
atoms to well below the temperature of quantum degeneracy
(Fermi
temperature $T_F$)
\cite{jila1,rice,ens,duke1,duke2,mit1,lens}, 
and the ratio $T/T_F=0.05$ has
been reached 
in the recent MIT experiment \cite{mit2}. One of the main
goals of these 
studies is achieving a transition to a superfluid
Cooper-paired
state. Trapped Fermi gases are very cold and dilute, with
temperatures
$T\alt 1$ $\mu$K and densities $n\sim 10^{13}$ cm$^{-3}$,
and for an attractive interspecies interaction (negative
$s$-wave
scattering length $a$) the most efficient will be the
superfluid $s$-wave
pairing between atoms of different components. However, 
the superfluid  transition temperature
$T_c$ is
exponentially small compared to the Fermi temperature $T_F$
and is beyond
experimental reach for ordinary small values
of $a$.  

At present, actively discussed ideas to circumvent this
difficulty rely on
superfluid pairing between fermionic atoms via a Feshbach
resonance \cite{holland,timmermans,griffin}. 
In the vicinity of the resonance the scattering length is
very large, being
negative on one side of the resonance and positive  on the
other side. 
On approach to the
resonance, the gas enters a strong-coupling regime. 
This occurs when the parameter $k_F|a|> 1$, with
$k_F=\sqrt{2mT_F}/\hbar$ being
the Fermi momentum and $m$ the atom mass.
Crossing the
resonance and making the scattering length positive,  the
formation of weakly
bound  dimers of two different fermions becomes
energetically favorable. 
Sufficiently far from resonance on the positive side, one
has a weakly interacting 
gas of these composite bosons and encounters the problem of
their Bose-Einstein 
condensation (BEC). This cross-over to the BEC
regime has been 
discussed in literature in the context of superconductivity
\cite{leggett,noz,randeria}.  

The BEC regime of the bosonic dimers is interesting from a
fundamental point
of view as it couples the problem of superfluidity in Fermi
gases to the
problem of molecular condensates. Here the most important
questions are the
stability of the condensate with regard to elastic
dimer-dimer 
interactions, and the decay of the gas due to relaxation of
the dimers to deep bound states. The relaxation occurs in 
dimer-dimer collisions and in collisions of dimers with
remaining 
unbound atoms. It is a crucial process as these dimers are  
diatomic molecules in the
highest rovibrational state. Several experiments show that
such molecules
consisting of bosonic $^{87}$Rb \cite{heinzen,rempe} and 
$^{133}$Cs atoms
\cite{rudi1}, or fermionic $^{40}$K atoms
with a scattering
length $a\sim 100$\AA \cite{jila2}, undergo a rapid
collisional
decay. On the other hand, recent observations 
\cite{salomon,rudi2,randy} indicate the existence of 
long-lived weakly bound Li$_2$ dimers at densities 
$\sim 10^{13}$ cm$^{-3}$.

In this Letter we present an exact solution for the
dimer-dimer elastic scattering, assuming that the
(positive) scattering length greatly exceeds the
characteristic radius of
interaction between atoms, $R_e$.
Then, as in the case of the 3-body problem with 
fermions (see \cite{efimov,petrov1} and references in
\cite{petrov1}),
the amplitude of elastic interaction is
determined by a single
parameter, the two-body scattering length $a$, and can 
be found in the zero-range approximation for the 
interatomic potential. Our findings 
lead to a positive dimer-dimer scattering length
$a_{dd}=0.6a$ \cite{strinati}. 
This is quite different from the assumption of earlier 
studies, $a_{dd}=2a$ \cite{randeria}, and has serious 
consequences for the low-temperature
behavior of the system.

We then discuss the collisional relaxation of the weakly 
bound dimers to deep bound states and show that it is 
{\it suppressed} due to {\it Fermi statistics} for the 
atoms. 
The binding energy of the dimers is
$\varepsilon_0=\hbar^2/ma^2$ 
and their size is close 
to $a$. The size of deep bound states is of the
order of $R_e\ll a$. 
Hence, the relaxation requires the
presence of at least three fermions at distances $\sim R_e$
from each other. 
As two of them are necessarily identical, due to the Pauli
exclusion 
principle the relaxation probability acquires a small factor 
proportional to a power of $(kR_e)$,
where $k\sim 1/a$ is a characteristic 
momentum of the atoms in the weakly bound molecular state. 
The inequality $a\gg R_e$ allows us to obtain the dependence
of the atom-dimer and dimer-dimer wavefunctions at short
interparticle distances on the two-body scattering length
$a$ and thus
to establish a strong decrease of the relaxation rate with
increasing $a$.
Our results show that the rate constant is
proportional to $a^{-s}$, where $s$ is close to $3$. 
This indicates a remarkable
collisional stability of the weakly bound bosonic dimers at
a 
large positive $a$, which is consistent with recent
observations at 
ENS~\cite{salomon}, Innsbruck~\cite{rudi2}, and
Rice~\cite{randy}. 
Note that the elastic collisional rate is proportional to
$a^2$, and for  
large $a$ we obtain a very large ratio of this rate to the
inelastic
rate of relaxation. 

The key idea behind these studies is twofold. 
First, a large ratio of the elastic to inelastic rate is
promising for 
evaporative cooling of the dimers, achieving their BEC, and
further cooling of
the Bose-condensed gas to very low temperatures.
Molecular BEC is a great challenge and many ongoing
studies are directed to
reaching this goal. Second, adiabatically 
crossing the Feshbach resonance and  reaching a negative
value of the 
scattering length $a$, the molecules are converted into
fermionic atoms
at extremely low temperatures $T\sim 0.01T_F$ which
can be
below the BCS transition temperature $T_c$ \cite{carr}.
Remarkably, 
at these temperatures the strong Pauli blocking of elastic
collisions provides the
collisionless regime for the atomic Fermi gas even for a 
large negative $a$. This opens up the possibility of 
identifying the BCS-paired state via 
the asymmetric expansion of the cloud released 
from a cylindrical trap~\cite{stringari}.

We start with the dimer-dimer elastic scattering which is a
4-body problem described by the Schr\"odinger equation
\begin{eqnarray}\label{4bodySchr}
&&-\left[\nabla_{{\bf
r}_1}^2+\nabla_{{\bf r}_2}^2+\nabla_{\bf
{R}}^2+mE/\hbar^2\right]\Psi
=-(m/\hbar^2)  \nonumber \\ 
&&\times\Big[ U(r_1)+U(r_2)+\sum_\pm U\left(({\bf r}_1+{\bf
r}_2\pm\sqrt{2}\,{\bf R})/2\right)\Big]\Psi.
\end{eqnarray}
Here ${\bf r}_1$ is the distance 
between two given distinguishable fermions,
${\bf r}_2$ is the distance between the other
two, ${\bf {R}}/\sqrt{2}$ is the distance
between the centers of mass of these pairs, and $U$ is the
interatomic potential. The total energy is
$\!E\!=\!-2\varepsilon_0\!+\!\varepsilon\!$, 
with $\varepsilon$ being the collision energy.  

The wavefunction $\Psi$ is symmetric with respect to the
permutation of the dimers and antisymmetric with respect to
permutations 
of identical fermions: 
\begin{eqnarray}\label{symmetry}
&&\Psi({\bf r}_1,{\bf r}_2,{\bf {R}})=\Psi({\bf
r}_2,{\bf r}_1,-{\bf {R}})  \\
&&=-\Psi\left(\frac{{\bf r_1}+{\bf r_2}\pm\sqrt{2}{\bf
{R}}}{2},\frac{{\bf r_1}+{\bf r_2}\mp\sqrt{2}{\bf
{R}}}{2},\frac{\pm({\bf r_1}-{\bf r_2})}{\sqrt{2}}\right).
\nonumber
\end{eqnarray}
At energies $\varepsilon\ll\varepsilon_0$ the 
scattering is dominated by the contribution of the $s$-wave
channel and can be analyzed from the solution of 
Eq.~(\ref{4bodySchr}) with $E=-2\varepsilon_0$. For large
${R}$ the corresponding wavefunction is given by 
\begin{equation}   \label{asymptote}
\Psi\approx\varphi_0(r_1)\varphi_0(r_2)(1-\sqrt{2}a_{dd}/R),
\end{equation}
where $a_{dd}$ is the dimer-dimer scattering length, and
\begin{equation}   \label{weakf}
\varphi_0(r)=(r\sqrt{2\pi a})^{-1}\exp{(-r/a)}
\end{equation}
is the wavefunction of a weakly bound dimer.

The characteristic de Broglie wavelength of atoms 
is of the order of $a$ and 
it greatly exceeds the radius $R_e$ of the interatomic
potential $U$.
Hence, for finding the scattering amplitude one can use the
well
known method of pseudopotentials, that is, to replace the
potential $U$ by a pseudopotential providing 
proper 
boundary conditions for $\Psi$ at vanishing distances
between two distinguishable fermions. 
For ${\bf r_1}\rightarrow 0$ the boundary condition reads:
\begin{equation}\label{boundary}
\Psi({\bf r}_1,{\bf r}_2,{\bf {R}})\rightarrow f({\bf
r}_2,{\bf
{R}})(1/4\pi r_1\,-1/4\pi
a),
\end{equation}
where the function $f({\bf r}_2,{\bf {R}})$ carries
the 
information about the second pair of particles when the 
first two sit on top of each other.
The boundary conditions for ${\bf r}_2\rightarrow 0$,
and for ${\bf r}_1+{\bf r}_2\pm\sqrt{2}{\bf R}\rightarrow 0$
are
readily written using Eqs.~(\ref{symmetry}).
The knowledge of the function $f$ is sufficient
for describing all of the four boundaries. Combining
Eqs.~(\ref{asymptote}) 
and (\ref{boundary}) one can deduce the value of $a_{dd}$
from the behavior of $f$ at large ${R}$. 
It is important that in contrast to $\Psi$, the function $f$ depends 
only on three variables: the absolute values of ${\bf r}_2$ and ${\bf
{R}}$, and the angle between them. 
In the following, we derive
and solve the equation for $f$.

In the pseudopotential approach the interaction
potential can be written as 
$U(r)=(4\pi\hbar^2a/m)\delta ({\bf r})(\partial/\partial r)
r$. 
Then, making use of Eq.(\ref{boundary}) we express the rhs
of
Eq.(\ref{4bodySchr}) in terms of the function $f$. 
The first term reads
$-mU(r_1)\Psi/\hbar^2=\delta ({\bf r_1})f({\bf r}_2,{\bf
{R}})$, and similarly for the other terms. The
resulting inhomogeneous Poisson equation can be solved by
inverting the differential operator. The Green
function of this operator is unique ($E=-2\varepsilon_0<0$), 
and $\Psi$ is expressed through $f$ as:
\begin{equation}\label{Psi}
\!\!\Psi(S)\!\!=\!\!\!\!\!\int\limits_{{\bf r}',{\bf
{R}}'}\!\!\!\!\bigl[\sum_{i=1,2}\!G(|S-S_i|)\!-\!\!\sum_\pm
G(|S\!-\!\!S_\pm|)\bigr]f({\bf r}'\!,{\bf {R}}'),
\end{equation}
where $S\!=\!\{{\bf r}_1,{\bf r}_2,{\bf
{R}}\},\;S_1\!=\!\{0,{\bf
r}',{\bf {R}}'\},\;S_2\!=\!\{{\bf r}',0,\!-{\bf
{R}}'\}$, $S_\pm\!\!=\!\!\{{\bf r}'\!/2\!\pm\!{\bf
{R}}'\!/\sqrt{2},{\bf
r}'\!/2\!\mp{\bf {R}}'\!/\sqrt{2},\!\mp{\bf
r}'\!/\!\sqrt{2}\}$, and the Green function is 
$G(x)=(2\pi)^{-9/2}(xa/\sqrt{2})^{-7/2}
K_{7/2}(\sqrt{2}\,x/a)$, with 
$K_{7/2}$ being the decaying Bessel function.

The equation for the function $f$ is obtained from
Eq.~(\ref{Psi})
by taking the limit $r_1\rightarrow 0$ and using
Eq.~(\ref{boundary}). 
The singular terms proportional to $1/r_1$
cancel each other automatically, and matching the regular
parts yields
\begin{eqnarray}\label{main}
&&\!\!\int\limits_{{\bf r}',{\bf
{R}}'}\!\!\Big\{G(|{\bar S}_1-S_1|)[f({\bf r}',{\bf {R}}')
-f({\bf r},{\bf {R}})]+\bigl[G(|{\bar S}_1-S_2|) \nonumber\\
&&\!\!-\sum_\pm\!
G(|{\bar S}_1\!\!-\!S_\pm|)\bigr]f({\bf r}'\!,{\bf {R}}')\!\Big\}\!
=\!(\sqrt{2}\!-\!1)f({\bf r},{\bf {R}})/4\pi a,
\end{eqnarray} 
where ${\bar S}_1=\{0,{\bf r},{\bf {R}}\}$. Note that the
only
length scale in Eq.~(\ref{main}) is $a$.
Solving this equation numerically we obtain the function $f$
for
all distances ${\bf R}$ and ${\bf r}$. According to 
Eqs.~(\ref{asymptote}), (\ref{weakf}) and (\ref{boundary}),
at 
$R\rightarrow\infty$ one has $f=(2/ra)\exp(-r/a)
(1-\sqrt{2}a_{dd}/R)$. Fitting our results 
with this expression we find with 2\% accuracy:
\begin{equation}   \label{add}
a_{dd}=0.6 a>0.
\end{equation}
 
Our calculations show the absence of 4-body weakly bound
states, and the behavior of $f$ at small ${R}$
suggests a soft-core repulsion between
molecules, with a range $\sim a$. The result of
Eq.~(\ref{add}) indicates the stability of molecular BEC with respect to
collapse. Compared to earlier studies which were assuming
$a_{dd}=2a$ \cite{randeria}, Eq.~(\ref{add}) gives almost twice as
small a sound velocity of the molecular condensate and a rate of 
elastic collisions smaller by an order of magnitude. 

The lifetime of the Bose gas of weakly bound dimers is
determined
by the process of their collisional relaxation into deep
bound 
states. The released binding energy 
of a deep state is of the order of $\hbar^2/mR_e^2$. 
It is transformed into the kinetic energy of particles in
the 
outgoing collisional channel and they escape 
from the trapped sample. We will establish a dependence of
the relaxation loss rate on the scattering length $a$,
without
going into a detailed analysis of the short-range behavior 
of the systems of three and four atoms. 
We employ a perturbative scheme assuming that the amplitude
of the
inelastic process of relaxation is much smaller than the
amplitude
of elastic scattering. 
Then the dependence of the relaxation rate on the scattering
length $a$ 
is related only to the $a$-dependence of the initial-state
wavefunction 
$\Psi$.

We first discuss the relaxation of weakly bound dimers to a
deep 
bound state in their collisions with atoms. 
This process occurs when all of the three atoms approach 
each other to distances $\sim R_e$. As at all interparticle 
distances $r\ll a$ the
three-body wavefunction  $\tilde \Psi$ is determined by the
Schr\"odinger equation with zero collision energy, it
depends on $a$ only through a normalization coefficient:
$\tilde \Psi=A(a)\tilde \psi$. 
In the region where $R_e\ll r\ll a$, the
$a$-independent function $\tilde \psi$ can be found in
the  zero-range approximation. Then the coefficient $A(a)$,
which determines the dependence of the relaxation rate on
$a$, is obtained by matching $\tilde \Psi$
with the solution at large interparticle  distances. 

This requires us to find $\tilde\Psi$ at 
distances $\sim a$, which we do using the zero-range 
approximation described in Ref. \cite{petrov1}. 
As in the case of four particles, we introduce the
corresponding function $\tilde f({\bf r})$. The correctly
symmetrized wavefunction $\tilde \Psi({\bf x},{\bf y})$ is
expressed
through $\tilde f$ as: 
\begin{equation}\label{Psi3body}
\tilde \Psi=\sum_\pm\!\int\limits_{\bf r}\! \pm {\tilde
G}\left(\!\sqrt{({\bf x}-{\bf r}/2)^2+({\bf
y}\mp\sqrt{3}{\bf r}/2)^2}\right)\tilde f({\bf
r}), 
\end{equation}
where ${\bf y}$ is the distance between identical
fermions, $\sqrt{3}{\bf x}/2$ is the distance
between their center of mass and the third atom,
and the Green function ${\tilde
G}(z)=(8\pi^2z^2a^2)^{-1}K_2(z/a)$. At a vanishing distance 
between two given distinguishable
fermions ${\bf r}_{\pm}=(\sqrt{3}{\bf x}\pm {\bf y})/2
\rightarrow 0$, the function $\tilde \Psi$
(\ref{Psi3body}) should have a correct asymptotic
behavior in analogy with Eq.~(\ref{boundary}). This gives
an equation for the function $\tilde f({\bf r})$: 
\begin{equation}\label{LE} 
\!\!\int\limits_{\bf r'}\!{\tilde G}\!\left(|{\bf
r}\!-\!{\bf r'}|\right)\![\tilde f({\bf
r'})\!-\!\tilde f({\bf r})]\!-\!{\tilde
G}\!\left(\!\!\sqrt{\!{\bf r}^2\!+\!{\bf r'}^2\!\!+\!{\bf
r}{\bf r'}}\!\right)\!\tilde f({\bf r'})\!\!=\!0.\!\!\!  
\end{equation} 

For small separations
between all three fermions, Eqs.~(\ref{LE}) and
(\ref{Psi3body}) give the same result as the method of
hyperspherical harmonics \cite{fedorov}: $\tilde
\Psi\!\approx\!
A(a)\rho^{\gamma}\Phi_{\gamma}(\Omega)$, where
$\Omega$ is a set of  hyperangles and the hyperradius
$\rho\!=\!\sqrt{x^2+y^2}$ satisfies the inequality $\rho\ll
a$.
The parameter $\gamma$ depends on the symmetry of  the
wavefunction and on the orbital angular
momentum of the atom-dimer motion. For the $s$-wave
atom-dimer scattering we have $\tilde f({\bf r})\!=\!\tilde
f(|{\bf r}|)$ and obtain
$\gamma\!=\!\gamma_1\!\approx\!0.1662$. 
At large atom-dimer distances $y$ we have   
$\Psi\approx \pm\varphi_0(r_{\pm})(1-a_{ad}/y)$,
where $\varphi_0$ is given by Eq.~(\ref{weakf}),
and $a_{ad}\!=\!1.2a$ is the atom-dimer scattering length
\cite{ter,petrov1}.
Solving Eqs.~(\ref{LE}) and (\ref{Psi3body}) at distances of
the order of $a$ and thus matching the two asymptotic
solutions, we find
$A(a)\!=\!const\times\!a^{-3/2-\gamma_1}$.
This result is explained extending the
wavefunction $\Psi$ to interparticle distances
$\sim\!a$ from the region of small distances. This gives
$\Psi\!\sim\!A(a)a^{\gamma_1}$, whereas being extended  to 
distances $\sim a$ from large
separations, the wavefunction is $\Psi\!\sim\!1/a^{3/2}$.
Equalizing the two expressions leads to the obtained
coefficient $A(a)$. 
The numerical constant is not needed as the absolute
value of the relaxation rate is determined by the 
contribution of interparticle distances $\sim\!R_e$, where
the 
zero-range approximation is not valid. It only gives a
correct 
dependence of the rate on $a$. 

As the relaxation rate constant $\alpha_{{\rm rel}}\propto
A^2(a)$,
we obtain $\alpha_{{\rm rel}}\propto
a^{-3-2\gamma_1}=a^{-3.33}$.
One thus sees that the relaxation rate due to dimer-atom
collisions
rapidly decreases with increasing two-body scattering length
$a$.

The obtained results can be easily generalized to the case
of the $s$-wave dimer-dimer scattering. Indeed, the
relaxation
process requires only three atoms to approach each other to
short distances. The fourth particle does not participate.
Let $\rho$ and $\Omega$ be the hyperradius and the set of
hyperangles of the system of three fermions and ${\bf
x}$ the distance between their center of mass and the
fourth atom. At $\rho\sim R_e$ and $|{\bf x}|\gg R_e$ the
four-particle wavefunction decomposes into
$\rho^{\gamma_2}\Phi_{\gamma_2}(\Omega)\eta({\bf x})$,
where the function $\eta({\bf x})$ describes the motion of
the
fourth particle. Averaging the relaxation probability over the motion
of the fourth particle makes the problem similar to the
relaxation in atom-dimer collisions, and we arrive at the same result
for the rate constant. However, there is a relaxation channel that
is more important in the limit of very large $a$. 
For the $s$-wave dimer-dimer scattering,
both the fourth particle and the atom bound to this particle
can
undergo $p$-wave scattering on the other dimer in such a way
that
their total angular momentum is equal to zero. This
corresponds to
an antisymmetric function $\tilde f({\bf r})$ in 
Eqs.~(\ref{Psi3body}) and (\ref{LE}) and leads to
$\gamma_2=-0.2273$. For the relaxation rate
we then find $\alpha_{{\rm rel}}\propto a^{-3-2\gamma_2}=a^{-2.55}$. 

Assuming that the short-range physics is characterized by
the length
scale $R_e$ and the energy scale $\hbar^2/mR_e^2$, 
we can restore the dimensions and write:
$\alpha_{{\rm rel}}= C (\hbar R_e/m)(R_e/a)^s$,
where $s\approx 2.55$ for the dimer-dimer collisional
relaxation,
and $s\approx 3.33$ for the relaxation in atom-dimer
collisions. 
The coefficient $C$ depends on a particular system and 
is different for dimer-dimer and atom-dimer collisions. 
Obviously, in the limit $R_e/a\rightarrow 0$ the dimer-dimer
relaxation should dominate over the atom-dimer one.

The much slower collisional decay of the weakly bound
dimers at larger $a$, following from our results, is 
consistent with recent observations for Li$_2$ 
molecules~\cite{salomon,rudi2,randy}, 
although the experimental data do not yet
provide the dependence of $\alpha_{{\rm rel}}$ on $a$.  
Nevertheless, the ENS result 
$\alpha_{{\rm rel}}\approx 2\times 10^{-13}$ cm$^3$/s for
the Li$_2$ loss rate at $a\approx 1500a_0$ \cite{salomon}, 
allows us to establish the coefficient $C\approx 20$. 

We emphasize that the remarkable stability of
such weakly bound dimers at $a\gg R_e$ is due to 
{\it Fermi statistics}.
Indeed, two identical fermions participating in the
relaxation process have very small relative momenta $k\sim
1/a$ and, 
hence, the process is suppressed compared to the case of a
dimerized gas of bosonic atoms.

The inelastic rate constant $\alpha_{{\rm rel}}$ is much
smaller than
the rate constant of elastic collisions $8\pi a_{dd}^2v_T$, where
$v_T$ is the 
thermal velocity. For the Li$_2$
weakly bound dimers at a temperature $T\sim 3\mu$K and 
$a\sim 1500$$a_0$, the corresponding ratio is of the order
of  
$10^{-4}$ or $10^{-5}$. This is nearly the same as in a cold
gas
of Rb atoms at densities $n\sim 10^{14}$ cm$^{-3}$. 
For Li$_2$ the values of both rate constants are larger, and
one can
expect even a higher  efficiency of evaporative cooling. It
should be possible
to reach BEC  of the dimers and cool the Bose-condensed gas
to temperatures of 
the order of its chemical potential. 

Then, converting the molecular BEC into fermionic atoms by 
adiabatically sweeping across the Feshbach resonance to the
negative side \cite{infinity}, one obtains the atomic Fermi
gas at extremely low temperatures
$T\sim 10^{-2}T_F$ which can be below the BCS
transition temperature $T_c$ \cite{carr}. At these
temperatures one has
a very strong Pauli blocking of elastic collisions. The
collisional rate
is suppressed as $15(T/T_F)^{2}$ \cite{duke2}, i.e. by a
factor of $10^3$,
and one expects the collisionless regime for the Fermi gas.      
This circumvents the main difficulty
for identifying the BCS-paired state via a free expansion of
the cloud released from 
the trap. This difficulty was met in experiments with
hydrodynamic Fermi clouds \cite{duke2}
which in cylindrical traps are characterized by the same
asymmetry of expansion
below and above $T_c$. Omitting small mean-field
effects, the collisionless cloud above $T_c$ expands
symmetrically,
whereas in the superfluid state the expansion becomes
significantly
asymmetric \cite{stringari}.

We acknowledge discussions with M.A. Baranov, C. Lobo,
and L.D. Carr. 
This work was supported by the Dutch Foundations NWO and
FOM, by 
INTAS, and by the Russian Foundation for Fundamental
Research. 
Laboratoire Kastler Brossel is a Unit\'e de Recherche de
l'Ecole Normale 
Sup\'erieure et de l'Universit\'e Paris 6, associ\'ee au
CNRS.


\begin{thebibliography}{99}
\bibitem{jila1} 
B. DeMarco and D.S. Jin, Science {\bf 285}, 1703 (1999); 
T. Loftus {\it et al}., Phys. Rev. Lett. {\bf 88}, 173201
(2002); M. Greiner {\it et al}., e-print cond-mat/0308519.
\bibitem{rice}
A.G. Truscott {\it et al}., Science {\bf 291}, 2570
(2001).
\bibitem{ens}
F. Schreck {\it et al}., Phys. Rev. Lett. {\bf 87},
080403 (2001).
\bibitem{duke1}
S.R. Granade {\it et al}., Phys. Rev. Lett. {\bf 88},
120405 (2002).
\bibitem{duke2} 
K.M. O'Hara {\it et al}., Science {\bf 298}, 2179 (2002).
\bibitem{mit1}
Z. Hadzibabic {\it et al}., Phys. Rev. Lett. {\bf 88},
160401 (2002); K. Dieckmann {\it et al}., {\it ibid.} 
{\bf 89}, 203201 (2002).
\bibitem{lens}
G. Roati {\it et al}., Phys. Rev. Lett. {\bf 89}, 150403
(2002); G. Modugno {\it et al}., Science {\bf 297}, 
2240 (2002).
\bibitem{mit2} 
Z. Hadzibabic {\it et al}., e-print cond-mat/0306050.
\bibitem{holland}
M. Holland {\it et al}., Phys. Rev. Lett. {\bf 87},
120406 (2001); J. Milstein {\it et al}., Phys. Rev. A 
{\bf 66}, 043604 (2002).
\bibitem{timmermans} 
E. Timmermans {\it et al}., Phys. Lett. A {\bf 285}, 228
(2001).
\bibitem{griffin}
Y. Ohashi and A. Griffin, Phys. Rev. Lett. {\bf 89}, 
130402 (2002). 
\bibitem{leggett}
A.J. Leggett, in {\it Modern Trends in the Theory of
Condensed Matter}, edited by A. Pekalski and J. Przystawa 
(Springer, Berlin, 1980).
\bibitem{noz}
P. Nozi\`eres and S. Schmitt-Rink, J. Low Temp. Phys. {\bf
59}, 195 (1985).
\bibitem{randeria}
See for review M. Randeria, in {\it Bose-Einstein
Condensation}, edited by A. Griffin, D.W. Snoke, and S.
Stringari (Cambridge University Press, Cambridge,1995).
\bibitem{heinzen}
D.J. Heinzen {\it et al}, Workshop on Cold Molecules,
Les Houches, 2002. 
\bibitem{rempe}
S. Durr {\it et al}., e-print cond-mat/0307440.
\bibitem{rudi1}
J. Herbig {\it et al}., submitted to Science.
\bibitem{jila2}
C.A. Regal {\it et al}., Nature {424}, 47 (2003). 
\bibitem{salomon}
J. Cubizolles {\it et al}., e-print cond-mat/0308018.
\bibitem{rudi2}
S. Jochim {\it et al}., e-print cond-mat/0308095.
\bibitem{randy}
K.E. Strecker, G.B. Partridge, and R.G. Hulet,
e-print cond-mat/0308318. 
\bibitem{efimov}
V.N. Efimov, Yad. Fiz. {\bf 12}, 1080 (1970) [Sov. J. Nucl.
Phys. {\bf 12},
 589 (1971)]; Nucl. Phys. A {\bf 210}, 157 (1973).  
\bibitem{petrov1}
D.S. Petrov, Phys. Rev. A {\bf 67}, 010703 (2003).
\bibitem{strinati} 
Diagrammatic approach used in P. Pieri and G.C. Strinati,
Phys. Rev. B 
{\bf 61}, 15370 (2000) gives $a_{dd}=0.75a$. However, 
this approach misses a number of diagrams which give a
contribution of the
same order of magnitude as those taken into account.
\bibitem{carr} 
L.D. Carr, G.V. Shlyapnikov, and Y. Castin, e-print
cond-mat/0308306.
\bibitem{stringari}
C. Menotti, P. Pedri, and S. Stringari, Phys. Rev. Lett.
{\bf 89},
250402 (2002). 
\bibitem{ter}
G.V. Skorniakov and K.A. Ter-Martirosian, Zh. Eksp. Teor.
Phys. {\bf 31}, 775 (1956)
[Sov. Phys. JETP {\bf 4}, 648 (1957)].
\bibitem{fedorov} 
See for review E. Nielsen, D.V. Fedorov, and A.S. Jensen, 
Phys. Rep. {\bf 347}, 374 (2000). 
\bibitem{infinity}
From a similar analysis as above, we find that for $a\rightarrow\infty$
the inelastic formation of deep bound states in a degenerate Fermi gas
occurs at a rate $\nu\sim (\hbar/mR_e^2)(k_F R_e)^{s+3}$. This rate is small
as $(k_F a)^s$ compared to the relaxation rate of weakly bound dimers.
\end{thebibliography}
\end{document}